# Suppression of ferromagnetism in rippled La$_{2/3}$Sr$_{1/3}$MnO$_3$ membrane with process-induced strain prepared by epitaxial lift-off technique

Kota Kanda[1,*], Ryuji Atsumi[1,*], Takamasa Usami[2,3], Takumi Yamazaki[4], Kohei Ueda[1,2], Takeshi Seki[4,5], Shigeki Miyasaka[6], Jobu Matsuno[1,2], and Junichi Shiogai[1,2,†]

[1]Department of Physics, Osaka University, Toyonaka, Osaka 560-0043, Japan

[2]Division of Spintronics Research Network, Institute for Open and Transdisciplinary Research Initiatives, Osaka University, Suita, Osaka 565-0871, Japan

[3] Center for Spintronics Research Network, Graduate School of Engineering Science, Osaka University, Toyonaka, Osaka 560-8531, Japan

[4]Institute for Materials Research, Tohoku University, Sendai, Miyagi 980-8577, Japan

[5]Center for Science and Innovation in Spintronics, Tohoku University, Sendai, Miyagi 980-8577, Japan

[6]Department of Material Science, Graduate School of Science, University of Hyogo, Ako, Hyogo 678-1297, Japan

[*]These authors equally contribute to this work.

[†]Corresponding authors: junichi.shiogai.sci@osaka-u.ac.jp

**ABSTRACT**

Transition metal oxides are a platform for exploring strain-engineered intriguing physical properties and developing spintronic or flexible electronic functionalities owing to strong coupling of spin, charge and lattice degrees of freedom. In this study, we exemplify the strain-engineered magnetism of $La_{2/3}Sr_{1/3}MnO_3$ in freestanding and rippled membrane forms without and with process-induced strain, respectively, prepared by epitaxial lift-off technique. We find that the deposition of Pt/Ti stressor suppresses the crack formation in the lift-off process and induces a ripple structure in the $La_{2/3}Sr_{1/3}MnO_3$ membrane. Laser micrograph and Raman spectroscopy show a ripple period of about 30 μm and a height of a few μm, where alternating convex and concave structures are subjected to tensile strain of 0.6% and compressive strain of 0.5%, respectively. While the freestanding $La_{2/3}Sr_{1/3}MnO_3$ membrane exhibits room-temperature ferromagnetism, the macroscopic magnetic transition temperature ($T_C$) of the rippled membrane is reduced by as large as 27%. Temperature-variable Kerr microscopy observation in the rippled membrane reveals that the spatial variation of $T_C$ to be approximately 4% of the macroscopic $T_C$, which coincides with the local strains at convex and concave structures. The large reduction of macroscopic $T_C$ in the rippled membrane may be ascribed to the lattice disorders due to strain gradient. Our demonstration of tuning ferromagnetism by the ripple structure validates the high potential of the process-induced strain in epitaxial lift-off technique and paves the way for strain-mediated emerging physical properties in various transition metal oxides.

**Main**

Strain often plays an important role for tuning physical properties in materials with strongly-localized electrons, where their spin and charge degrees of freedom are coupled to lattices such as bond lengths and angles, as well as coordination symmetry [1]. In such strongly-correlated systems, tuning their magnetic or electronic properties with external and/or internal strains are particularly attractive because it offers a promising platform for magnetostrictive energy harvesting [2][3], strain gauge application [4], and development of other spintronic or flexible electronic functionalities [5][6][7]. With the aim of modulation of magnetic properties in such functional materials by strain engineering, an effective approach for strain application is a matter of interest.

One representative example is an epitaxial strain in thin-film growth by selecting proper lattice-mismatched substrates. In the thin-film approach, manganite perovskites have been intensively studied: modulation of their electron correlation and magnetic ground states has been demonstrated by introducing in-plane tensile or compressive epitaxial strains [8]. However, the large and continuous control of strain is difficult in this approach because the maximum applicable strain is limited to about 3% by formation of dislocations [9] and the lattice constant of the available substrate is discrete. Another example is a single or few atomic-layer-thick exfoliated flakes of van der Waals (vdW) magnets prepared on flexible substrate or in suspended form. In these sample configurations, the magnetic properties can be well controlled by stretching the base substances as exemplified in the $Fe_3GeTe_2$ and CrSBr flakes [10][11]. Further, the process-induced strain, in which depositing thin-film stressors transfers their residual strain on the underlayer, has been utilized for lateral imprinting of strain-sensitive physical properties. Although this approach has been originally developed for high-performance complementary metal-oxide-semiconductor (CMOS) applications [12], it has been recently applied to the vdW materials flakes [13][14] owing to its high compatibility with freestanding nature of the vdW materials and conventional lithography technique [15].

Recently, the epitaxial lift-off technique [16] using a water-soluble $Sr_3Al_2O_6$ sacrificial layer has attracted considerable attentions [17][18] for synthesis of freestanding membranes of various transition metal oxides [19][20][21][22][23] and their heterostructure [17][24]. One extreme characteristic of the freestanding membrane is the mechanical robustness against various forms of strain such as biaxial or uniaxial form with a large stretching up to 8 % [21], and gradient or bending form [25]. To date, strain-mediated magnetic phase transition has been demonstrated in various functional magnets with membrane form such as a stretched membrane of manganite [21] or rippled membranes of Heusler compound [26][27]. While the high utility of the process-induced strain is also expected in freestanding membrane of the strongly-correlated materials, modulation of their magnetic properties by this approach has not been explicitly demonstrated so far.

Here, we apply the process-induced strain to the membrane of prototypical ferromagnetic transition metal oxide La$_{2/3}$Sr$_{1/3}$MnO$_3$, prepared by epitaxial lift-off technique and find that the Pt/Ti thin-film stressor plays a crucial role as a supporter for great suppression of crack formation and a thin-film stressor for inducing a ripple structure. The induced strain in rippled membrane drives a magnetic phase transition from ferromagnet to paramagnet at room temperature. Our experimental results suggest a high potential of process-induced strain not only for conventional CMOS technology or vdW materials but also for exploring emergent physical properties of freestanding membrane of the strongly-correlated materials.

The freestanding membrane of La$_{2/3}$Sr$_{1/2}$MnO$_3$ was obtained by epitaxial lift-off technique from a thin-film heterostructure of 60-nm-thick La$_{2/3}$Sr$_{1/2}$MnO$_3$ layer and 17-nm-thick Sr$_3$Al$_2$O$_6$ sacrificial layer prepared on SrTiO$_3$(001) substrate by pulsed-laser deposition. As depicted in Fig. 1(a), the La$_{2/3}$Sr$_{1/2}$MnO$_3$ film (bulk lattice constant $a_{\text{LSMO}}$ = 3.894 Å [28]) is subjected to the tensile strain from the Sr$_3$Al$_2$O$_6$ underlayer ($a_{\text{SAO}}/4$ =

3.961 Å). By dipping the heterostructure in distilled water, the $Sr_3Al_2O_6$ sacrificial layer was dissolved and the tensile strain in the $La_{2/3}Sr_{1/2}MnO_3$ layer was fully released as previously reported [23]. Concomitantly, a significant number of cracks with their typical lateral size of ~ 10 μm were randomly distributed in the released $La_{2/3}Sr_{1/2}MnO_3$ membrane as seen in the micrograph [Fig. 1(b)].

This situation is drastically changed when depositing 100-nm-thick Pt/10-nm-thick Ti cap layer on the $La_{2/3}Sr_{1/2}MnO_3$ layer prior to etching the sacrificial layer. [Fig. 1(c)]. Figure 1(d) shows the top view micrograph of the Pt/Ti/$La_{2/3}Sr_{1/2}MnO_3$ membrane. Here, the micrograph was taken from the $La_{2/3}Sr_{1/2}MnO_3$ layer. After the etching, the formation of the cracks was strongly suppressed, and a large continuous rippled membrane with its lateral size of the order of mm was obtained [Fig. 1(e)]. We consider that the strong adhesive property of the Ti layer between the top Pt layer and the bottom oxide layer plays a crucial role for the $La_{2/3}Sr_{1/2}MnO_3$ layer to be released on the mm scale [29]. The analysis of the x-ray diffraction patterns (see Supplementary Figure S1) indicates that the Pt layer is subjected to 0.75% compressive strain in the plane and 2% tensile strain out of the plane. According to our previous report [23], on the other hand, the $La_{2/3}Sr_{1/2}MnO_3$ layer is subjected to 2.1% tensile strain in the plane at the interface with the $Sr_3Al_2O_6$ layer, and fully relaxed at around 16 nm apart from the interface. We speculate that the transferring the compressive strain of the Pt stressor on the relaxed top surface of the La$_{2/3}$Sr$_{1/2}$MnO$_3$ layer and relaxation of tensile strain in the strained bottom surface of the La$_{2/3}$Sr$_{1/2}$MnO$_3$ layer drive the formation of the ripple structure during the releasing process.

To deepen the understanding of detailed structure for the rippled membrane, laser microscopy and Raman spectroscopy were carried out for height and strain profiles, respectively. Figure 2(a) shows a two-dimensional height profile on a typical area of the rippled membrane. Apparently, the rippled membrane possesses convex and concave structures with a period of tens μm in the plane and a height of a few μm [see also one-dimensional height profile along white dashed line in bottom image of Fig. 2(a)]. Here, the strain $\varepsilon$ in convex and concave in the rippled membrane composed of three layers is estimated by the relation [30][31],

$$\varepsilon = \frac{1}{R} \frac{\sum_{i=1}^{3} E_i t_i \left[ \left( \sum_{i=1}^{3} t_i \right) - \frac{t_i}{2} \right]}{\sum_{i=1}^{3} E_i t_i},$$

where $R$ is a radius of curvature when we regard the bent La$_{2/3}$Sr$_{1/2}$MnO$_3$ layer surface as a circular sector [see the inset of Fig. 2(c)], $E_i$ and $t_i$ represent Young's modulus [32] and thickness of $i$-th layer, respectively. The averaged values of $\varepsilon$ are calculated to be + 0.6% and −0.5% at convex and concave structures, respectively, along the white line in Fig. 2(a). Figure 2(b) shows Raman spectra obtained at convex (purple circles in the inset) and concave (red) of the ripple structure after background signal is subtracted. Note that the spot diameter of the present spectrometer is 15 μm, which is small enough to capture the properties specific to the convex and concave structures. As can be seen, two broad Raman peaks were observed around 540 and 650 cm$^{-1}$, which we assigned to the $A_g$ (Jahn-Teller mode) and $B_{2g}$ (breathing mode) phonons, respectively [33]. The characteristic Raman spectrum corresponding to the Jahn-Teller mode obtained at convex and concave structures do not show a detectable difference, indicating that the impact of the ripple structure is insignificant on the out-of-phase vibration of Mn-O-Mn bonding of MnO$_6$ octahedrons. Figure 2(c) shows the relation of Raman shifts obtained by $B_{2g}$ phonons [red and purple vertical segments in Fig. 2(b)] and strain estimated by line profile [Fig. 2(a)] for convex (purple circle) and concave (red circle) as well as that in La$_{0.7}$Sr$_{0.3}$MnO$_3$ powder reported in previous work [33]. In contrast to the insensitive $A_g$ peak, we observe the slight shift of the $B_{2g}$ peak of breathing mode: the characteristic wave number for the convex (concave) structure is shifted lower (higher). The breathing mode corresponds to the in-phase vibration of Mn-O-Mn bonding of MnO$_6$ octahedrons, which is directly related to the average length of Mn-O-Mn bonding. For previous powder experiments [33], the compressive strain by chemical doping exhibits the red shift with respect to the non-doped La$_{0.7}$Sr$_{0.3}$MnO$_3$, which is qualitatively consistent with our estimation shown in Fig. 2(c). From these observations, we conclude that the bending strain in the convex and concave structure modulates the Mn-Mn bonding distance more effectively rather than the Jahn-Teller strain.

The impact of the ripple structure on magnetic properties was first characterized by the magnetic field ($H$) dependence of magnetization [$M(H)$ curve] of the freestanding and rippled La$_{2/3}$Sr$_{1/3}$MnO$_3$ membranes when $H$ was applied in the sample plane. Figures 3(a) and 3(b) show those $M(H)$ curves measured at $T = 300$ K. The fully relaxed freestanding membrane exhibits ferromagnetism with a good squareness, being consistent with the room-temperature ferromagnetism of La$_{2/3}$Sr$_{1/3}$MnO$_3$ [34] and our previous report [23]. In stark contrast, the rippled membrane does not show detectable magnetization at $T = 300$ K. Figures 3(c) and 3(d) show the set of $M(H)$ curves measured at $T = 5$ K. The $M(H)$ curves for the freestanding and rippled membranes fully saturate in the magnetic field range of 0.6 T and 0.5 T, respectively. The obtained value of saturation magnetization is 3.42 $\mu_B$/Mn for the freestanding membrane [Fig. 3(c)], close to the ideal value 3.67 $\mu_B$/Mn ($\mu_B$: Bohr magneton) under the assumption of the mixed valence of Mn$^{3+}$ and Mn$^{4+}$ ions in the stoichiometric La$_{2/3}$Sr$_{1/3}$MnO$_3$. This consistency also suggests the fully-relaxed nature of the freestanding membrane. On the other hand, the saturation magnetization is only 2.02 $\mu_B$/Mn for the rippled membrane as shown in Fig. 3(d).

Such sharp differences of the magnetization between the freestanding and rippled membranes are more clearly seen in temperature ($T$) dependence of magnetization [$M(T)$ curve] shown in Fig. 3(e). As expected by development of the ferromagnetism at $T$ = 300 K for the freestanding membrane [Fig. 3(a)], the $M(T)$ curves exhibit a high ferromagnetic transition temperature $T_C$ around $T$ = 370 K, well above room temperature. On the other hand, $T_C$ for the rippled membrane is strongly suppressed, where the $M(T)$ starts to develop around 270 K and reaches about 2 $\mu_B$/Mn at the lowest temperature. By comparing to the fully magnetized state in the freestanding membrane, it is suggested that the local structural variation *i.e.,* the tensile and compressive strain in the convex and concave structure, respectively, reduces $T_C$ by 27% and saturation magnetization in the rippled membrane.

To clarify the correlation between ripple structure and magnetism, we performed Kerr microscopy measurement with descending temperature. Figure 4(a) shows the raw image of reflection contrast in the rippled membrane. The bright contrast corresponds to alternating convex (purple arrows) and concave (red arrows) structures. The reflection from the area between convex and concave structures was not clearly detected because the $La_{2/3}Sr_{1/3}MnO_3$ surface is tilted from the normal to the sample stage. Although the convex and concave structures show almost the same brightness, the height profile was judged by defocusing while moving the sample stage along the *z*-axis. To extract only the contribution sensitive to the magnetization, we calculate a subtraction of the raw image taken at $\mu_0 H = -0.1$ T from that at $\mu_0 H = +0.1$ T. Figure 4(b) shows a differential image taken at $T = 290$ K, where no bright contrast is detected because the ferromagnetism is not well developed at this temperature. At $T = 280$ K, weak contrast starts to emerge along both convex and concave structures [Fig. 4(c)]. Below $T = 270$ K, the white lines in the concave are slightly brighter than those in the convex structure [Fig. 4(d)], and both contrasts become comparable at $T = 260$ K [Fig. 4(e)]. Such temperature variation of brightness is linked to the development of ferromagnetism in the $M(T)$ curve around $T = 270$ K. From these results, we conclude that the spatial variation of local $T_C$ by local strain at the convex and concave structures is approximately 10 K, which corresponds to 4% with respect to macroscopic $T_C$ of 270 K.

Finally, we quantitatively discuss the relation between local structure and magnetization. For $ABO_3$-type perovskites, the strength of *p-d* hybridization and thus, $T_C$ depends on both *B*-O-*B* bond angle $\theta_{B\text{-}O\text{-}B}$ and bond length $d_{B\text{-}O}$ (where $B$ = Mn in manganites) [35]. Judging from the $A_g$ mode insensitive to the ripple structure as shown in Fig. 2, we can assume that Mn-O-Mn bond angle is independent on the tensile or compressive strain in the present ripple structure. Based on the tight binding approximation, the *d-p* hybridization $t_{p\text{-}d}$ between Mn and O atoms is proportional to $1/d_{\text{Mn-O}}^{7/2}$ (Ref. 35). Since the Mn ions are well separated each other, the effective *d-d* interaction $t_{d\text{-}d}$ via the O ion is the square of $t_{p\text{-}d}$ (Ref. 36). Therefore, from the naive expectation of $T_\text{C}$ being proportional to $t_{d\text{-}d}$ and the assumption of $\theta_{\text{Mn-O-Mn}}$ being constant, the modulation of $T_\text{C}$ by strain between concave and convex structures ($\Delta\varepsilon \sim 1.1\%$) is estimated as $\Delta T_\text{C}/T_\text{C} \sim 7\Delta\varepsilon \sim 8\%$. This difference is in the same order as the spatial variation of local $T_\text{C}$ revealed in the results of the Kerr micrography (Fig. 4). In contrast, the reduction of macroscopic $T_\text{C}$ in the rippled membrane from the freestanding membrane is about 27%, much larger than that from our estimation. One plausible explanation of this discrepancy is the role of strain gradient in the rippled membrane, which is regarded as a quenched disorder [37] from the averaged structure or freestanding form. According to the Monte Carlo simulation performed for the optimally doped ferromagnetic manganites [38][39][40], it has been demonstrated that the quenched disorder enhances electron-phonon coupling, which suppresses the electron hopping probability and induces metal-to-insulator transition. Because of the strong coupling between electrical transport and magnetism in manganites, the enhanced quenched disorder is also related to the suppression of ferromagnetic order in the present rippled membrane. Further characterization of microscopic structure and numerical analysis of strain distribution are needed for more quantitative discussion.

In summary, we showed that the use of Pt/Ti layer as a stressor can suppress the crack formation and apply the process-induced strain into the La$_{2/3}$Sr$_{1/3}$MnO$_3$ membrane prepared by epitaxial lift-off technique, which resulted in the formation of ripple structure. The induced strain at the convex and concave structures in the rippled membrane effectively affects the Mn-O bonding length, which allows us to clearly modulate magnetic phases at room temperature. Considering the compatibility of process-induced strain to conventional lithography technique, our proof-of-concept study in application of this approach to the manganite membrane can be extended to artificial lateral patterning of the strain and strain gradient on two-dimensional membranes of various strongly-correlated materials.

Acknowledgments

The authors acknowledge Kohei Hamaya at Osaka University for Kerr microscopy, and Makoto Kohda at Tohoku University for laser microscopy, and Jason K. Kawasaki at University of Wisconsin-Madison for stimulating discussion. This work was partly performed under the GIMRT Program of the Institute for Materials Research, Tohoku University (Proposal No. 202212-CRKEQ-0007). The Raman experiment was performed using an NRS-3100T spectrometer (JASCO Co., Ltd.) with the help of Takashi Harada at the Research Center for Solar Energy Chemistry, Graduate School of Engineering Science, Osaka University. This work was supported by JST, PRESTO Grant No. JPMJPR21A8, and JSPS, KAKENHI Grant Nos. JP22K18894 and JP23K26379, and Iketani Science and Technology Foundation, and Tanikawa Foundation.

States in Models for Manganites: The Origin of the Colossal Magnetoresistance Effect. Phys. Rev. Lett. **98**, 127202 (2007).

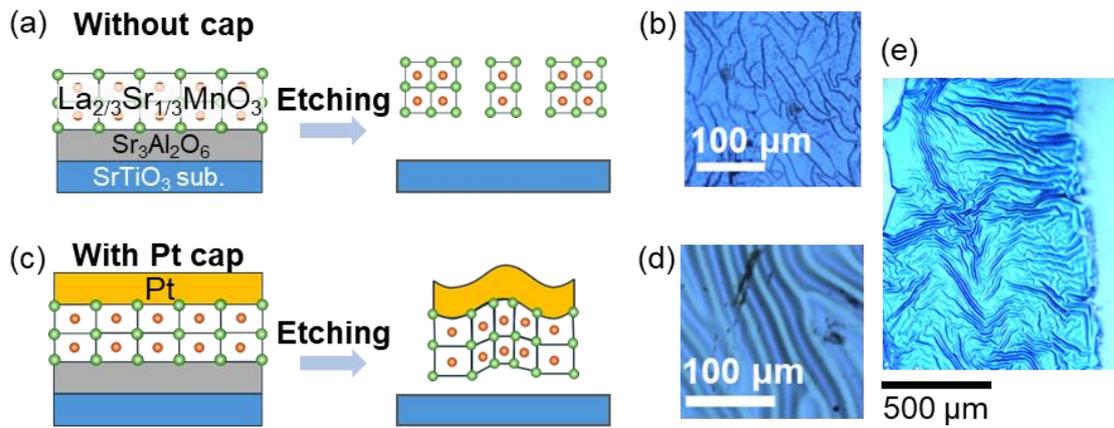

**FIG. 1.** (a) Procedure of epitaxial lift-off technique of the $La_{2/3}Sr_{1/2}MnO_3/Sr_3Al_2O_6$ heterostructure accompanied with epitaxial strain relaxation and crack formation. (b) Top-view micrograph of the freestanding $La_{2/3}Sr_{1/2}MnO_3$ membrane. (c) Formation of ripple structure in lift-off process for the Pt/Ti capped $La_{2/3}Sr_{1/2}MnO_3$ membrane. (d) Top-view micrograph of the rippled Pt/Ti/$La_{2/3}Sr_{1/2}MnO_3$ membrane taken from the $La_{2/3}Sr_{1/2}MnO_3$ face. (e) A wide view of (d) showing mm-scale crack-free membrane with ripple structure.

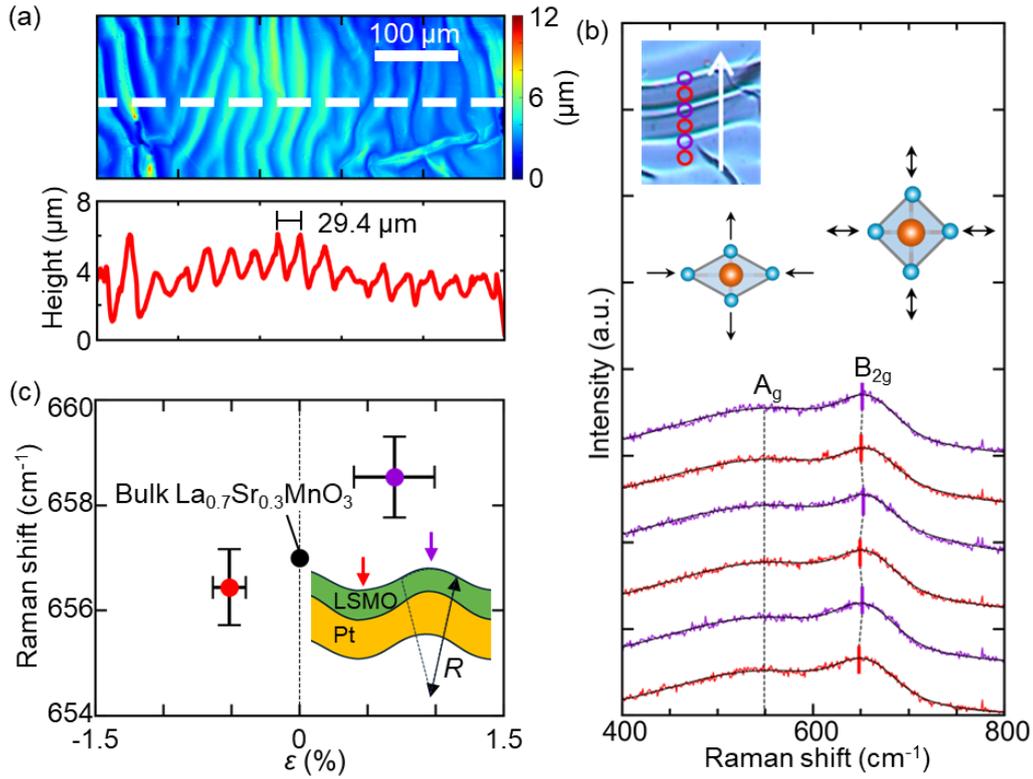

**FIG. 2.** (a) Laser micrograph of the rippled La$_{2/3}$Sr$_{1/2}$MnO$_3$ membrane. (Top) Color plot of areal height profile and (bottom) line profile along the white dashed line. (b) Raman spectrum measured at concave (red) and convex (purple) structures. For clarity, background is subtracted. Solid black lines indicate analysis of Gaussian fitting and vertical red and purple segments around $B_{2g}$ peak indicate extracted peak positions. (Inset) Top view photograph of measurement area. (c) Averaged values of the Raman shift for concave (red) and convex (purple) structures as well as that of powder La$_{0.7}$Sr$_{0.3}$MnO$_3$ sample reported previously (black) [32]. Vertical and horizontal error bars correspond to variation in cross-sectional curvatures and Raman shifts among investigated points, respectively. (Inset) Schematic of cross-sectional view of ripple structures.

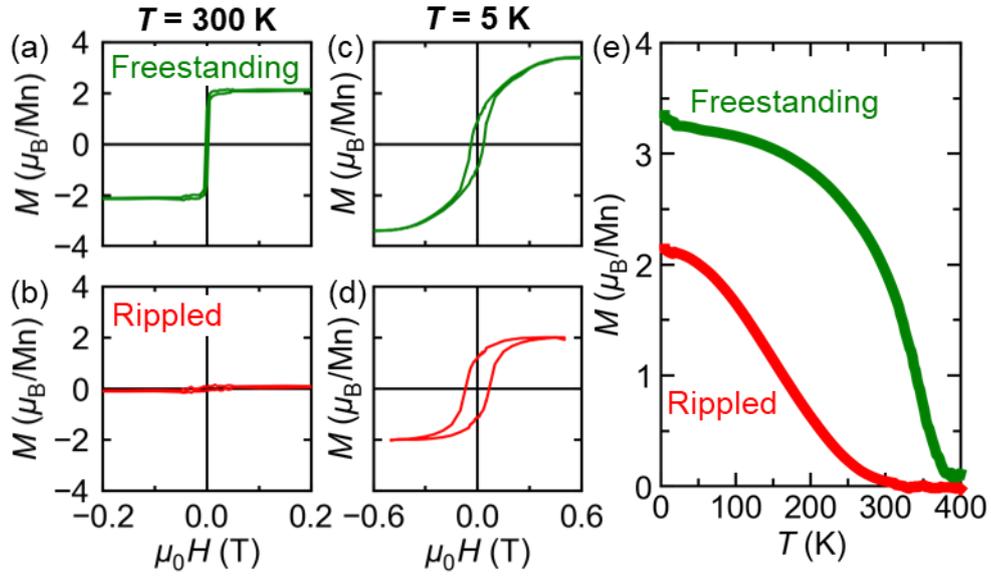

**FIG. 3.** Magnetic hysteresis loops for (a) freestanding and (b) rippled La$_{2/3}$Sr$_{1/3}$MnO$_3$ membranes measured at $T = 300$ K and those for (c) freestanding and (d) rippled membranes measured at $T = 5$ K. (e) Temperature dependence of magnetization $M(T)$ for freestanding (green) and rippled (red) membranes. Measurements were performed at an in-plane field of $\mu_0 H = 1.5$ T for the freestanding membrane and at $\mu_0 H = 0.5$ T for the rippled membrane.

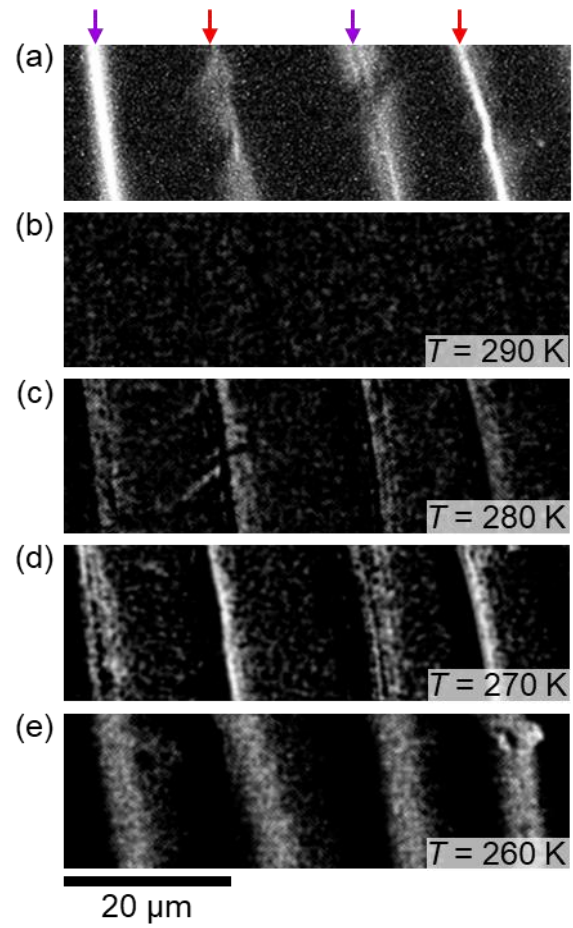

**FIG. 4.** (a) Reflectivity image obtained by Kerr microscopy. (b)-(e) Contrast difference of reflectivity images between at $\mu_0 H = +0.1$ T and $-0.1$ T, obtained at $T =$ (b) 290, (c) 280, (d) 270, and (e) 260 K. Purple and red arrows indicate convex and concave structures, respectively.